\documentclass[10pt, conference, compsocconf]{IEEEtran}
\usepackage{listings}
\let\labelindent\relax
\usepackage{enumitem}
\usepackage{algorithmicx}
\usepackage{algpseudocode}
\usepackage{color,soul}
\sethlcolor{white}
\usepackage[hyphens]{url}
\usepackage{hyperref}

\ifCLASSINFOpdf
  \usepackage[pdftex]{graphicx}
  \graphicspath{{../pdf/}{../jpeg/}}
  \DeclareGraphicsExtensions{.pdf,.jpeg,.png}
\else
\fi
\begin{document}
%
\title{Serverless Computing: Behind the Scenes of Major Platforms}


\author{\IEEEauthorblockN{Daniel Kelly}
\IEEEauthorblockA{School of Computer Science\\National University of\\
Ireland, Galway (NUIG),\\
Galway, Ireland.\\
Email: d.kelly69@nuigalway.ie}
\and
\IEEEauthorblockN{Frank G Glavin}
\IEEEauthorblockA{School of Computer Science\\National University of\\
Ireland, Galway (NUIG),\\
Galway, Ireland.\\
Email: frank.glavin@nuigalway.ie}
\and
\IEEEauthorblockN{Enda Barrett}
\IEEEauthorblockA{School of Computer Science\\National University of\\
Ireland, Galway (NUIG),\\
Galway, Ireland.\\
Email: enda.barrett@nuigalway.ie}}


%


\maketitle

\begin{abstract}
Serverless computing offers an event driven pay-as-you-go framework for application development. A key selling point is the concept of no back-end server management, allowing developers to focus on application functionality. This is achieved through severe abstraction of the underlying architecture the functions run on. We examine the underlying architecture and report on the performance of serverless functions and how they are effected by certain factors such as memory allocation and interference caused by load induced by other users on the platform. Specifically, we focus on the serverless offerings \hl{of the four largest platforms; AWS Lambda, Google Cloud Functions, Microsoft Azure Functions and IBM Cloud Functions}. In this paper, we observe and contrast between these platforms in their approach to the common issue of ``cold starts'', we devise a means to unveil the underlying architecture serverless functions execute on and we investigate the effects of interference from load on the platform over the time span of one month.

\end{abstract}


%
\IEEEpeerreviewmaketitle

\section{Introduction}
Serverless computing is an application deployment architecture that aims to provide pay-as-you go event driven functionality. Applications are developed as per each desired process and the event that invokes it. Serverless function platforms provide the infrastructure to deploy code for execution across their cloud and define the event processing logic that prompts the functions to run using the model: \emph{event, trigger, and action}. Serverless computing promotes the idea that application development is abstracted from the underlying infrastructure and that there is no need for a dedicated team to manage software in-house, sparking the term ``NoOps'' (no operations). Serverless computing abstracts back-end management from users, allowing only minimal access to some basic parameters such as function memory allocation and function runtime timeout. Functions execute on the platform's traditional Infrastructure as a Service (IaaS) virtual machine offerings, however the provision of such virtual machines (VM) is managed by the platform in response to function invocation and not by the developer. Unlike IaaS, you do not pay for the uptime and resources consumed by this ``execution'' VM but rather for the run time of each function, hence the name Function as a Service (FaaS).
Due to this abstraction enacted by the platform providers, it is not necessarily known what kind of infrastructure functions may be executed on. This paper aims to look behind the scenes of serverless functions and offer an up-to-date insight on the topology of \hl{the four largest platforms; AWS Lambda (AWS), Google Cloud Functions (Google), Microsoft Azure Functions (Azure) and IBM Cloud Functions (IBM)}, as well as outline performance metrics and potential effects on performance.

We designed a testing system to expose the underlying infrastructure that serverless functions run on. We gather the CPU specifications and usage statistics in order to provide this overview of the resources provisioned. To this end, a means of identifying VMs was established to create meaningful profiles of VM configurations. From this, we gained an insight into their topology and observed the major differences between the platforms. Users cannot directly request the specification of these execution VMs, only the memory allocation each function has access to and, as such, document how this configuration can effect the specification of the VM deployed. We further investigate how this memory allocation affects function performance by recording various timing statistics and benchmarking metrics associated with function execution such as: function runtime, ``cold start'' time (initialization lag), CPU utilization and disk I/O throughput. Lastly, given \hl{that} all the functions for a given platform are executed from the same cloud space (split into regions), the effect of potential interference from load on the platform was investigated by running our test system over the course of one month in order to detect any anomalies in runtime performance.

\section{Related Work}
There are countless blogs posted by serverless application developers that give a brief overview of the performance of serverless functions. \cite{CuiYan2017, CuiYan2017a, ShilkovMikhail2019, ByrroRenato2019, VoijtaRobert2016}. The focus of these is often on the issue of ``cold starts'' and programming language choice for function development. Given how quickly these articles are produced, the results often seem at odds with each other as time goes on. For example, Voijta \cite{VoijtaRobert2016} concluded that memory allocation of a function \hl{does not affect} cold start time, whereas Cui \cite{CuiYan2017a} concluded that it did one year later. In academia, a number of researchers have attempted to assess the performance of various serverless platforms. Hendrickson \emph{et al.} \cite{Hendrickson2016} created an open-source serverless platform using the \emph{Lambda Programming Model} (develop functions that respond to events), which achieved \hl{lower latency} at low loads and was better at bursts of traffic compared to AWS Elastic Beanstalk\footnote{\url{https://aws.amazon.com/elasticbeanstalk/}}. The authors also proposed a \hl{benchmarking} tool (LambdaBench) for the Lambda Programming Model. McGrath \emph{et al.} \cite{McGrath2017} analysed the performance of AWS Lambda, Google Cloud Functions, Microsoft Azure, and Apache OpenWhisk against their own prototype serverless platform. They made observations on the scaling and cold start latency of the platforms, finding AWS and Google to be best-in-class at the time, however the testing setup may not have been optimal for recording cold start times i.e. not allowing for enough time to pass where a cold start would be encountered. Llyod \emph{et al.} \cite{Lloyd2018} expand on the cold start issue by defining them as: \emph{provider cold} (first request to the cloud provider), \emph{VM cold} (first request to a VM within that cloud), and \emph{container cold} (first request to a container within that VM), as well as voicing the need for a standard means of benchmarking serverless functions inspired by the suggestions of Aderaldo \emph{et al.} \cite{Aderaldo2017} on microservices. Mohanty \emph{et al.} \cite{Mohanty2018} investigate the popular performance metrics used by other researchers with their study focusing on open source frameworks such as OpenFaas, Kubeless and Fission. There were no comparisons to the commercial offerings in this work. Hellerstein \emph{et al.} \cite{Hellerstein2018} provide a ``devil's advocate'' opinion on serverless architecture, listing limitations such as: limited function lifetime, I/O bottlenecks, no specialised hardware, how FaaS stymies distributed computing because there is no network addressability of serverless functions, two functions can work together ``serverlessly'' only by passing data through slow and expensive storage and how FaaS discourages open source service innovation since most popular open source software cannot not run at the same scale as current commercial serverless offerings. Wang \emph{et al.} \cite{Wang2018} performed an analysis of the topology of the serverless platforms as well as their performance. This work has inspired some of the methodology of the experiments conducted in this paper.

Despite the related work completed in the field, we have identified gaps in the knowledge that our work aims to fill. Namely, we investigate potential interference effects on performance caused by load on the cloud platform by other users over the course of one month. This has resulted in the generation of a data set containing over \hl{500,000 function calls between the four platforms} that we have published along with our benchmarking tools\footnote{\url{https://github.com/psykodan/Serverless-Computing-Data-Set}}. This work is performed on the current state of AWS Lambda, Google Cloud Functions, Microsoft Azure Functions and IBM Cloud Functions in 2020. Development is rapid and updated insight into their performance are required as previous findings become outdated.

\section{Methodology}
The typical case for invocation of serverless functions is through the use of REpresentational State Transfer (REST) APIs. For our tests, an observer virtual machine was deployed on each platform's traditional cloud computing offering. They were deployed on the same availability zone as the serverless functions in order to best nullify network latency. This observer's role was to invoke the functions via their API, record the request and response times and finally push the results to an external MongoDB database. The observer script was written in Python and used the \emph{requests} package to trigger each function via its URL. The observer script was executed in two scenarios:
\begin{enumerate}
\item Two sequential function invocations for the definitive measurement of a cold and warm function start. 
\item Fifty concurrent function invocations to prompt scaling and put increased workload on the function platform. 
\end{enumerate}
These scenarios were run every hour for one month.
The serverless functions execute a number of routines that perform measurements on the specifications of the function's runtime environment. Such measurements include: function runtime, VM uptime, total available memory on VM, disk I/O throughput, CPU information and unique identifiers for both function and VM instances. Much of these results were read from the Linux proc filesystem (procfs) where available.

\begin{figure}[!t]
\centering
\includegraphics[width=3in]{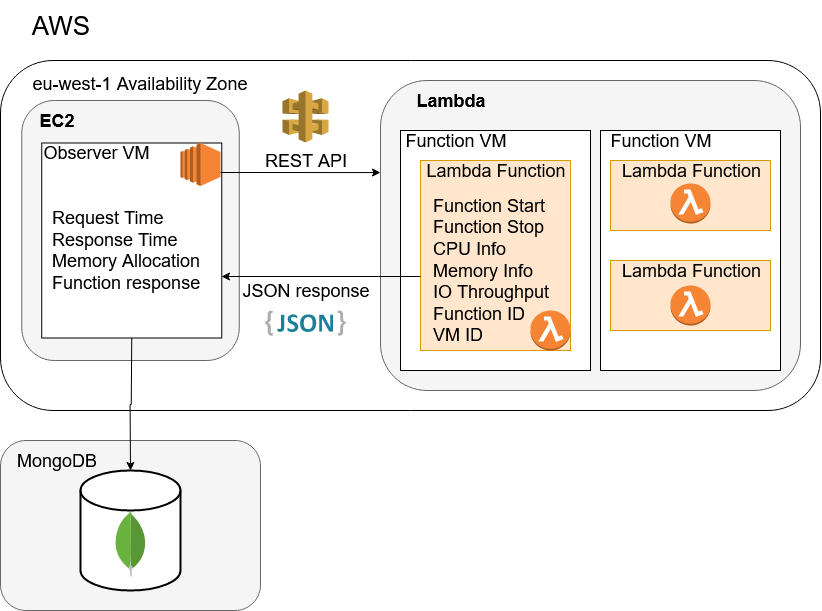}
\caption{AWS Experimental Setup}
\label{fig1}
\end{figure}

The results of these routines were collated into a JSON response that was sent back to the observer where it could append extra information such as: memory allocation of the function, total runtime of the function (request to response) and the start lag time (cold start time Section \ref{coldstart}), \hl{before storing the results} to the database. This process is illustrated above in Figure \ref{fig1} for AWS Lambda, a near identical setup is used for \hl{the other platforms} except using their equivalent cloud offerings. 

\section{Platform Architecture}
In order to understand the architecture of \hl{the four platforms}, we consulted the official documentation for each platform in addition to running our aforementioned tests. Table \ref{table1} shows the measurements that we recorded using the function for gathering information on the platform architecture.
\begin{table*}[!t]
\renewcommand{\arraystretch}{1.1}
\caption{Recorded Attributes of Execution VM}
\label{table1}
\centering
\begin{tabular}{|c||l|}
\hline
\multicolumn{2}{|c|}{\textbf{Identification}}\\
\hline
VM ID &A unique identifier for each execution VM. The method of measurement varies for each platform.\\
\hline
Function ID & A unique identifier for each function instance.\\
\hline
Previous Function IDs & A collection of Function IDs that have executed on a particular VM.\\
\hline
\multicolumn{2}{|c|}{\textbf{System Information}}\\
\hline
CPU Speed & The speed in MHz of the CPU.\\
\hline
CPU Times & The amount of time in milliseconds a host has spent in user, system and idle time.\\
\hline
CPU Model & What CPU is used on the host.\\
\hline
Uptime & The timestamp the VM booted.\\
\hline
Total Memory & Total memory configured to the VM\\
\hline
Memory & The function memory allocation.\\
\hline
\end{tabular}
\end{table*}



\subsection{Overview}
\textbf{AWS, Azure and IBM:} The proc file system exposes information about the execution VM that the functions run on and not just the function container. From our measurements, we found that the average \hl{VMs configured were: 

AWS - Intel Xeon E5-2670 v2 processor @ 2.5 GHz with memory that varied based on the memory allocation of the functions executed on it such that is was slightly greater than the function's memory allocation

Azure - Intel Xeon E5-2673 v4 @ 2.3 GHz with 2GB of memory

IBM - Intel Xeon E5-2683 v4 @ 2.1 GHz with 16GB of memory}

\textbf{Google Cloud Functions:} Functions run in isolation on Googles proprietary hypervisor. This obscures the data we would have gathered from the proc file system to show mostly nothing. From our test, however, we could conclude that the average VM was configured with a Intel Xeon Skylake processor @ 2 GHz \hl{with 2GB of memory.

In contrast to AWS, Google and Azure simply allocate all their execution VMs with the maximum memory requirement for the largest function configuration of 2048MB and IBM allocate a larger memory allocation perhaps to handle more function containers per VM.}

\subsection{Unique VM Identifier}
Uniquely identifying the VM used to execute the function allows us to gain a picture of the infrastructure each platform is using and can help to answer questions about scaling, function isolation, disk access etc. Work carried out by Wang \emph{et al.} \cite{Wang2018} determined that for AWS Lambda functions, the VM can be identified via the \texttt{/proc/self/cgroup} file, where an entry contains a unique identifier ``sandbox-root-'' followed by some random characters. \hl{For Azure, they proposed the use of an environment variable called ``WEBSITE\_INSTANCE\_ID''. However at the time of our experiments no such variable could be found. So for both Azure and IBM, the \texttt{/proc/machineid} file was used to identify VMs}. A means to uniquely identify a Google Cloud Function's execution VMs was not established. Due to the isolation Google's proprietary hypervisor imposes on the functions, access to the information normally stored in the proc file system is abstracted to show nothing, therefore an \hl{entry that identify a machine} could not be found. A heuristic to identify a VM by assuming they all have unique boot times by Lloyd \emph{et al.} \cite{Lloyd2018} was also disproved by Wang \emph{et al.} \cite{Wang2018}, deeming it unreliable. \hl{In order to gather meaningful data for comparison to the other platforms}, we propose a method of identifying function containers by writing some \emph{Unique Identifier} (UID) to a file into the \texttt{/tmp} folder. It was found that this file could be read by other functions executing \hl{in the same container}. This was used, in conjunction with the boot time information as a sanity check, to ensure that the UID is indeed allocated to one \hl{container}.

\subsection{VM Topology}
\label{topology}
The topology was examined from \hl{our data set of 563,276}  function invocations. \hl{This exposed 156856 VMs on AWS, 1392 VMs on Azure, 114 VMs on IBM and 21271 unique function containers on Google.
It was stated by Wang \emph{et al.} }\cite{Wang2018}\hl{, and confirmed ourselves through correspondence with Google employees, that Google Cloud Functions execute on their proprietary hypervisor that abstracts as much info about the execution VM as possible. However, we were successful in identifying function containers and could infer the CPU model information from the measurements we took. Two fields that were not hidden in the \texttt{/proc/cpuinfo} file were the CPU model id and the speed.} CPU models were determined using the model number entry in the \texttt{/proc/cpuinfo} file, which gives a decimal number that can be converted to hexadecimal and then cross referenced against the CPUID Signature table found in the Intel Architectures Software Developer's Manual, Volume 4 Chapter 2 \cite{Intel}.\hl{The CPU models found on each platform are collated in Table} \ref{cpuid}. \hl{All CPUs were a member of the Intel Xeon product line with IBM boasting the latest versions of such, comprising of mainly version 3 and 4 models. AWS has a more homogeneous CPU configuration than the other platforms, opting for 99.93\% of them being Intel Xeon E5-2670 v2.

We also measured the total memory configured to the VM.} In AWS, we observed that this value varied depending on the memory allocation of the function itself. A mostly consistent mapping was observed, as shown in Table \ref{table2}, except for two outliers in the 1024MB functions, which ran on a VM with 1717196kB of memory*. \hl{The other platforms had a constant memory configuration with Azure and Google allocating 2GB and IBM allocating 16GB.}
\begin{table}[!t]
\renewcommand{\arraystretch}{1.0}
\caption{AWS Lambda Function Memory to VM Total Memory}
\label{table2}
\centering
\begin{tabular}{|c|c|}
\hline
Function Memory (MB) & VM Total Memory (MB)\\
\hline
128 & 192.484\\
\hline
256 & 331.740\\
\hline
512 & 633.804\\
\hline
1024 & 1190.860\\
\hline
*1024 & 1717.196\\
\hline
2048 & 3230.668\\
\hline
\end{tabular}
\end{table}

\begin{table}[!t]
\renewcommand{\arraystretch}{1.0}
\caption{\hl{CPU Identification}}
\label{cpuid}
\centering
\begin{tabular}{|c|c|c|c|c|}
\hline
Platform & Model ID & Speed  & Model Name & Prevalence\\
&(decimal)&(MHz)&&(\%)\\
\hline
AWS&&&&	\\
\hline
&62 & 2500&	Xeon E5-2670 v2 &	99.93\\

\hline
&62&	3000&	Xeon E5-2690 v2 &	0.07\\

\hline
Google	&&&&\\
\hline
&45&	2600&	Xeon E5-2670&	24.12\\

\hline
&45&	3300&	Xeon E5-1660 &	0.02\\

\hline
&62	&2500	&Xeon E5-2670 v2 &	0.14\\

\hline
&63&	2300&	Xeon E5-2680 v3 &	4.79\\

\hline
&79&	2200&	Xeon E5-2650 v4 	&11.25\\

\hline
&85&	2000&	Xeon (Skylake)*	&53.11\\
\hline
&85&	2200&	Xeon (Skylake)*&	6.54\\
\hline
IBM	&&&&\\		
\hline
&85&	2300&	Xeon Gold6140 &	19\\

\hline
&79&	2100&	Xeon E5-2683 v4 &	42.68\\

\hline
&79&	2600&	Xeon E5-2690 v4 &18.83\\

\hline
&79&	2200&	Xeon E5-2650 v4 &2.6\\

\hline
&63&	2600&	Xeon E5-2690 v3 &	9.79\\

\hline
&85	&2100	&Xeon Gold6130 &	7.05\\

\hline
&63&	2000&	Xeon E5-2683 v3 &	0.04\\

\hline
Azure&&&&	\\		
\hline
&79&	2300&	Xeon E5-2673 v4 &	68.68\\

\hline
&63&	2400&	Xeon E5-2673 v3 &	22.07\\

\hline
&85&	2600&	Xeon Platinum8171M &	9.24\\

\hline
\end{tabular}
\footnotesize{* Specific model could not be determined}\\
\end{table}

\section{Function Performance}
\label{perf}
The extent to which a developer can customize the configuration of a serverless function is quite limited. For AWS you can alter the memory allocation by increments of 64kB from 128MB to 3096MB and choose the timeout from a range of 1 second to 5 minutes.\hl{ IBM has memory allocation in increments of 32kB from 128MB to 2048MB and a timeout of 1 second to 10 minutes.} Google allows you to choose from a predefined set of five memory allocations: 128MB, 256MB, 512MB, 1024MB and 2048MB. \hl{Finally, Azure does not allow the developer to configure their memory allocation opting instead for auto-scaling memory. }For our investigation of function performance, we used a series of measurements, as detailed in Table \ref{table4}, on functions with memory allocations 128MB, 256MB, 512MB, 1024MB and 2048MB \hl{for AWS, Google and IBM. For Azure, the same tests were run without the granularity of five memory allocations i.e. only one function that was allowed to auto-scale accordingly.} 
\begin{table*}[!t]
\renewcommand{\arraystretch}{1.0}
\caption{Measurements of Function Performance}
\label{table4}
\centering
\begin{tabular}{|c||l|}
\hline
Total Runtime & Measured from time API invokes the function to the time a response is received.\\
\hline
Function Runtime & The time taken for a function to execute its tasks not including initialization time of the function.\\
\hline
Start Lag & Function initialization time. Measured from request time to main method start time.\\
\hline
CPU Utilization & The number of primes a function can compute in 1000 ms.\\
\hline
Disk I/O & The I/O throughput of a function.\\
\hline
Number of VMs & The number of execution VMs created to handle scaling.\\
\hline
\end{tabular}
\end{table*}

\subsection{Cold Start Latency}
\label{coldstart}
\begin{figure}[!t]
\centering
\includegraphics[width=3.5in]{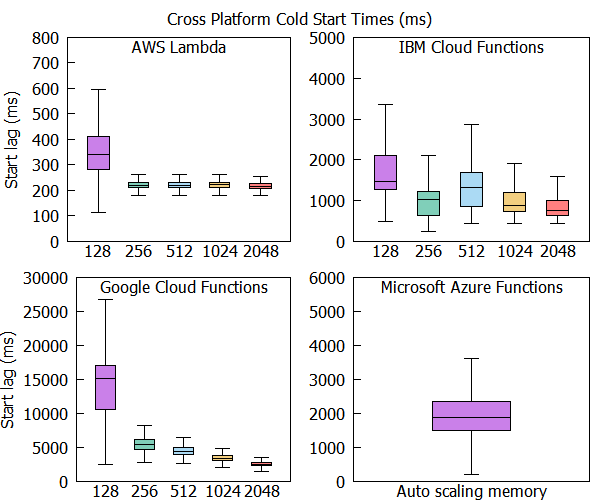}
\caption{\hl{Cross Platform Cold Start (request time to function start)}}
\label{cold}
\end{figure}

One of the greatest problems facing serverless computing is the infamous cold start. Cold starts have been the subject of research papers and blogs \cite{ShilkovMikhail2019,ShilkovMikhail2018,ByrroRenato2019,CuiYan2017,Lin2019,Baldini2017} and continue to be a major source of doubt for those considering a serverless based application. They are the notable delay incurred when invoking a function for the first time. This is caused by the need for the platform to spin up a container that has all the required resources for a function to run. We measured the time taken from the function request to the time a function's main method began executing. We gathered these times for functions that were the first to execute on a new \hl{function container}. This was determined by a file written to the \texttt{/tmp} directory that contained a log of all the IDs from functions that had previously executed in that container. \hl{The \texttt{/tmp} directory is ephemeral storage that lasts the lifespan of its function container.} If the log was empty, it was a new container. We sampled across a time span of one month, running the tests every hour. This resulted in over \hl{170,000 function invocations on AWS, Google and IBM and over 36,000 on Azure (a fifth of the number of invocations since no separate tests for each memory allocation) }for our analysis. Results are shown in Figure \ref{cold}.

\textbf{AWS:} \hl{Of 175,477 function invocations executed on AWS 156,847 were cold starts.} This would suggest that AWS may not prioritise the reuse of old containers. Of the five memory allocations chosen for the test function, the 128MB had the slowest average cold start time at \hl{346.73 ms. A point of interest is the near identical average values for the 256, 512, 1024 and 2048MB functions at 221ms $\pm$3ms.}  It may be possible that AWS gives more precedence to these latter memory allocations.

\textbf{Google:} \hl{Of the 175,357 function invocations executed on Google 21,253 were cold starts.} There is high container reuse on Google Cloud Functions, some containers were hosting over a thousand function executions. A more expected stepping in results for each memory allocation's average cold start time was observed. These values were considerably higher than AWS with averages for 128, 256, 512, 1024 and 2048MB being \hl{14465.52ms, 5722.33ms, 4681.37ms, 3689.48ms and 2865.49ms }respectively. 

\hl{
\textbf{IBM:}  Of 176,266 function invocations executed on IBM 37,820 were cold starts. Similar container reuse to Google is observed although with much shorter cold starts: 2990.55ms, 1076.60ms, 1310.43ms, 1319.05ms and 915.49ms for the respective memory allocations. The results do not show a consistent trend which is discussed in Section }\ref{inter}.

\hl{
\textbf{Azure:} Of 36,176 function invocations executed on Azure 1392 were cold starts. The average cold start was 1997.63ms. The low number of cold starts is likely to do with there being only one function being invoked rather than five thus allowing for greater container reuse.

From these results, we can see the different tactics employed by the different platforms. AWS opts for less container reuse (shorter lifespan) which results in more cold starts, although the cold start times were considerably shorter than the others. Google and IBM reused containers much more thus minimizing cold starts. This is especially useful for Google whose cold starts were considerably higher.}
\subsection{CPU Utilization}
Given a 1000ms time limit, check as many numbers as possible for whether it is prime. The method we use to check for prime numbers is trial division. 
For some number n
across a set \begin{equation}1 < a \leq \sqrt{n}\end{equation}
n is prime if \begin{equation}\gcd{n,a} \geq 1\end{equation}


\begin{algorithmic}
\Function{$isPrime$}{$n$}
    \State $start\gets 2$
    \State $limit\gets \sqrt{n}$
\While{$start\leq limit$}
    \If{$n<1 \pmod{start}$}
        \State \Return $False$
    \Else
        \State $start\gets start+1$
    \EndIf
\EndWhile

\Return $True$
\EndFunction
\end{algorithmic}

\begin{figure}[!t]
\centering
\includegraphics[width=3.5in]{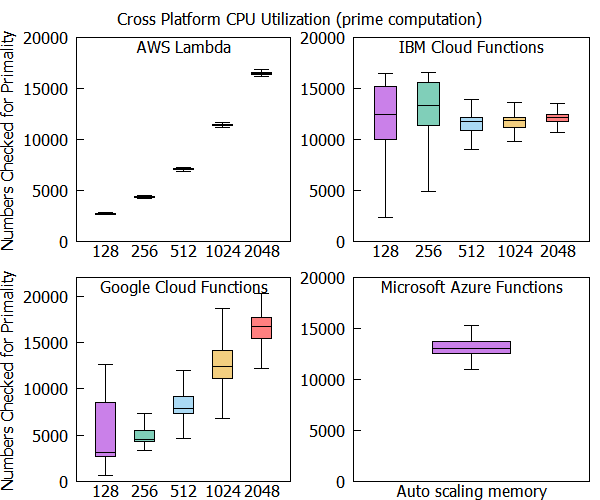}
\caption{\hl{Cross Platform no. numbers checked for primality}}
\label{cpu}
\end{figure}

This is a laborious way to calculate primes with the sole intention of using up CPU resources. The results are based on the volume of numbers checked for primality one-by-one (Figure \ref{cpu}).

\textbf{AWS:} The volume of numbers the function could check for primality increased as the memory allocation increased. AWS specifies that the CPU power is distributed using time slicing, with larger memory allocations receiving longer time slices. The method of time slicing yields consistent access to the CPU as demonstrated by our results i.e. the number of numbers checked had a low range of variation. 

\textbf{Google:} Average values were similar to that of AWS. However, each memory allocation had quite a wide range of values. This is likely due to the larger variety of CPUs used by the execution VMs resulting in less consistent results.

\hl{
\textbf{IBM:}  Average values were similar across all memory allocations with less spread of values in the higher allocations. These values are comparable to the 1024MB functions in AWS and Google. There is considerable spread in the lower memory allocations. However we believe this is due to interference on the platform (discussed in Section }\ref{inter}).

\hl{
\textbf{Azure:} Again, a similar volume of primes were computed to the other platforms

The similarity in the results suggest that the differences in topology (Table} \ref{cpuid}\hl{) are not necessarily a factor governing a function's CPU utilization. The greater factor is likely that of the method each platform employs for CPU resource allocation. Time-slicing is the most suitable method of allocating resources as one can then use a more homogeneous back-end for function execution, reducing complexity. The algorithm for time-slicing will be the deciding factor in a function's CPU utilization.
}

\subsection{Disk I/O Throughput}

\begin{figure}[!t]
\centering
\includegraphics[width=3.5in]{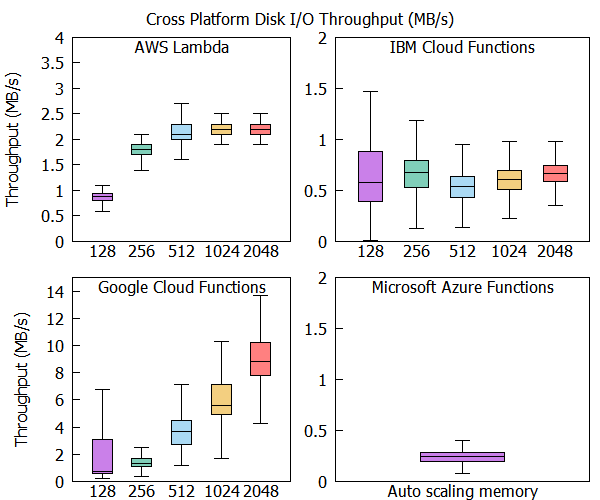}
\caption{\hl{Cross Platform Disk I/O Throughput}}
\label{disk}
\end{figure}

Serverless functions are more commonly associated with reading and writing to a proprietary storage solution: AWS Lamba $\rightarrow$ Amazon Simple Storage Service (S3) and Google Cloud Functions $\rightarrow$  Google Filestore. However, both offerings also allow for the reading and writing of temporary files to disk on the execution VM. Our disk throughput tests utilize the Linux ``dd'' command to read and write blocks of size 512KB 1000 times and outputs disk throughput in MB/s (Figure \ref{disk}). 

\textbf{AWS:} Throughput increased as memory allocation increased sharply after 128MB. However, it tapered towards the larger memory allocations, never getting above 3MB/s. Given the more homogeneous topology of the execution VMs observed, we can infer that that this is the limit that is approached as the function is allocated more CPU time slices with a greater memory allocation.

\textbf{Google:} A more varied range was recorded with an increasing throughput for greater memory allocations. This is likely related to the much greater number of\hl{ CPU} configurations (Table \ref{cpuid}). Another recording of note is how much greater the values are to those from AWS. It is stated in Google Cloud Functions' documentation that these temporary files are actually stored in memory rather than on disk\footnote{\url{https://cloud.google.com/appengine/docs/standard/python3/using-temp-files}} which may be resulting in the greater throughput. 

\hl{
\textbf{IBM:} Average values were consistent and approximately 0.6MB/s mark for each memory allocation. This, along with the previous test further suggests that IBM Functions are not affected by memory allocation.

\textbf{Azure:} The throughput recorded for Azure was the lowest of all the platforms, averaging under 0.5MB/s.

The disk throughput is important for more complex functions that depend on temporary memory access. As functions are billed on execution time, lower throughput may result in increased run time.
}

\subsection{Interference Effect on Performance}
\label{inter}
\begin{figure}[!t]
\centering
\includegraphics[width=3.5in]{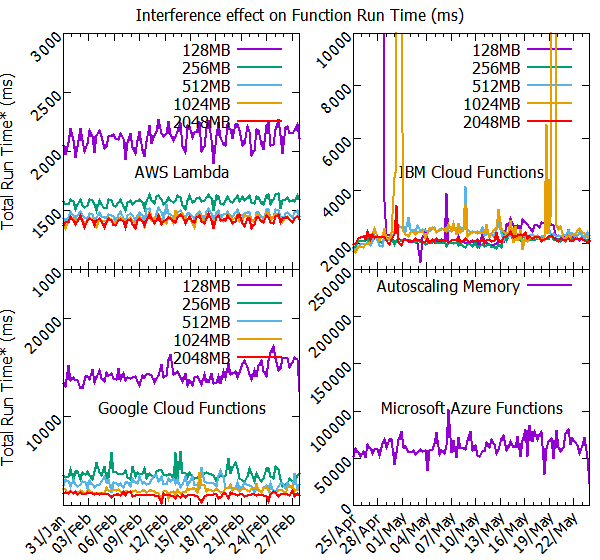}
\caption{\hl{Function run times over one month *Smoothed hourly values}}
\label{run}
\end{figure}

\begin{figure}[!t]
\centering
\includegraphics[width=3.5in]{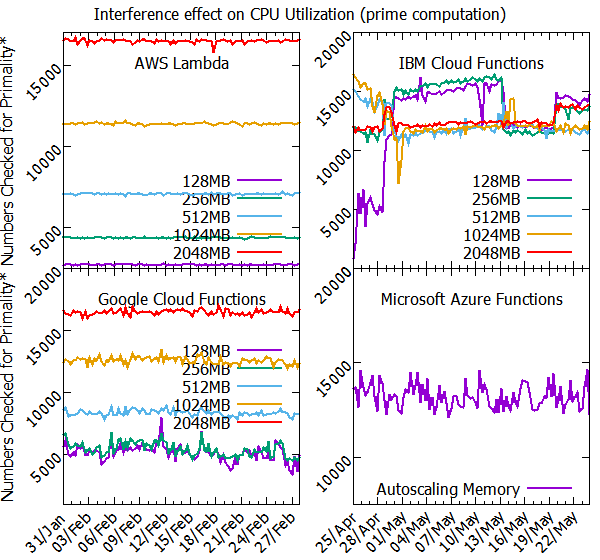}
\caption{\hl{Function CPU Utilization (prime computation) over one month *Smoothed hourly values}}
\label{cpulong}
\end{figure}

\begin{figure}[!t]
\centering
\includegraphics[width=3.5in]{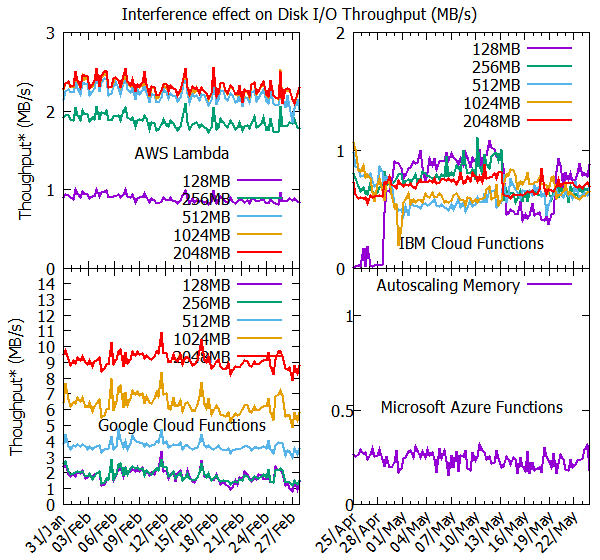}
\caption{\hl{Function Disk I/O Throughput over one month *Smoothed hourly values}}
\label{disklong}
\end{figure}

Serverless platforms employ a similar strategy on how functions are executed. A function runs within a function container on an execution VM within a predefined region of the serverless platorm's cloud. The execution VMs are isolated from other individual users \cite{Wang2018}. \hl{However, different users' VMs can exist in the same region potentially leading to interference effects. Other events such as wide scale updates, system outages etc. may also interfere with function execution. We ran our tests over the course of one month to investigate the potential effect of any interference that causes anomalies or predicable fluctuations in performance. We examined the effect on function run time, CPU utilization and disk I/O throughput. The results are a smoothed graph of data points taken from each hourly run of our test functions using gnuplot's ``acspline''}\footnote{\url{http://gnuplot.sourceforge.net/docs\_4.2/node125.html}} for run time (Figure \ref{run}), CPU utilization (Figure \ref{cpulong}) and disk I/O throughput (Figure \ref{disklong}).

\hl{
\textbf{AWS:} Running in region ``eu-west-1'', we observed minimal variance in values for CPU utilization. However, a sinusoidal pattern is visible in the results of the effect on run time and disk I/O throughput. The pattern becomes clearer in the higher memory allocations. We believe this is due to higher memory allocations having greater access to the CPU (via time slicing) and, as such, will execute in a more consistent manner, therefore amplifying the visual effect of interference when it occurs. The pattern has peaks at 12:00pm and troughs at 12:00am each day, which correlates to higher use during the day than the night. This may be evidence to support the potential effect of regular load on the system.}

\hl{
\textbf{Google:} Running in region ``us-central1-a'', more erratic values were observed. However, consistent peaks and troughs are visible, similar to AWS. Unlike AWS, Google Cloud did not have a consistent topology for its execution VMs (Section }\ref{topology}\hl{), this may be the cause of the varied array of results. AWS Lambda functions predominantly ran on the same CPU configurations, meaning their results were more consistent, allowing us to see clearer interference patterns.

\textbf{IBM:} Running in region ``Dallas'', clear signs of strain on the platform were observed with large dips in performance occurring at the beginning and near the end of our month of testing. There were a number of issues affecting IBM's cloud during this time according to their status history page\footnote{\url{https://cloud.ibm.com/status?selected=history}}, which may be the culprit. Another possibility is that IBM's serverless platform is not capable of consistent performance like its competitors leading to anomalies like the ones captured by our tests. 

\textbf{Azure:} Running in region ``Central US'', results similar to Google were observed being erratic in amplitude but periodic in time peaks and troughs in its performance.

Our month long analysis demonstrates that there is interference on each platform and we propose potential sources for such interference. As serverless functions are billed per 100ms, we can see that the fluctuations observed from our tests would have an effect on the final cost. As well as cost there is also considerations to be made for more critical functions requiring consistent performance. 
}

\section{Conclusion and Future Work}
\hl{
We performed an in-depth analysis of the four biggest commercial serverless platforms' underlying topology and the effects on performance for differing memory allocations. We found that AWS has a different approach to function container reuse that we theorise to be a design choice to address scaling and cold start issues. We determined that memory allocation largely effects the performance of serverless functions on these platforms and must be a serious consideration when developing applications. Finally, we performed a month long observation on function performance, creating a large data set of function benchmarks and uncovering potential interference effects due to increased load on the platform during the day and strain caused by maintenance on the platform. We believe that the contributions of this paper can help unveil the often mysterious inner workings of serverless platforms, which seem to have based their business model on keeping the developer, not just as far away from the back-end as possible, but in the dark about it. We have produced a produced data set that contains a lot of data on the running of serverless function on the four largest platforms. It will benefit to any future work on application benchmarking, optimization via machine learning or cybersecurity.}

It is becoming the norm to evaluate serverless functions using similar methods as displayed in this paper \cite{Wang2018, CuiYan2017a, ShilkovMikhail2018} as well as real world work loads. Proposals for a standardised benchmarking system are certainly not differing from this trend \cite{Kim2019}. A vendor agnostic, multi-language, benchmarking tool is a necessary step to encourage a greater uptake of serverless as the architecture of choice for application development. As is stands, a developer must use intuition for memory allocation rather than an empirical evaluation. Potentially greater control can be given to the developer for the configuration of functions. Certain aspects of the serverless architecture such as total isolation of functions could potentially be tweaked for fringe case applications making it a more robust architecture.





%

\bibliographystyle{IEEEtran}
\bibliography{serverless.bib}
\end{document}